# A Physically Based Compact I–V Model for Monolayer TMDC Channel MOSFET and DMFET Biosensor

Ehsanur Rahman[1$**], Abir Shadman[1$], Imtiaz Ahmed[1], Saeed Uz Zaman Khan[1] and Quazi D. M. Khosru[1]

[1] Department of Electrical and Electronic Engineering, Bangladesh University of Engineering and Technology, Dhaka-1205
**To whom correspondence should be addressed. Email: ehsaneeebuet@gmail.com
$ These two authors have equal contribution to this work

**Abstract: In this work, a compact transport model has been developed for monolayer transition metal dichalcogenide channel MOSFET. The analytical model solves the Poisson's equation for the inversion charge density to get the electrostatic potential in the channel. Current is then calculated by solving the drift-diffusion equation. The model makes gradual channel approximation to simplify the solution procedure. The appropriate density of states obtained from the first principle DFT simulation has been considered to keep the model physically accurate for monolayer transition metal dichalcogenide channel FET. The outcome of the model has been benchmarked against both experimental and numerical quantum simulation results with the help of few fitting parameters. Using the compact model, detailed output and transfer characteristics of monolayer WSe₂ FET have been studied, and various performance parameters have been determined. The study confirms excellent ON and OFF state performances of monolayer WSe₂ FET which could be viable for the next generation high-speed, low power applications. Also, the proposed model has been extended to study the operation of a biosensor. A monolayer MoS₂ channel based dielectric modulated FET is investigated using the compact model for detection of a biomolecule in a dry environment.**

## 1. Introduction

Analytical and compact modeling can give better insight into the operation of a device. For monolayer Transition Metal Dichalcogenides (TMDCs), classical transport models will not be appropriate because of the presence of high degree of confinement. In 2012, Jiménez [1] presented a physics-based model of the surface potential and drain current for monolayer TMDC FET. The work took the 2D density of states of the monolayer TMDC and its impact on the quantum capacitance into account and modeled the surface potential. The author further developed an expression for the drain current considering the drift-diffusion mechanism. The analytical expressions of surface potential and drain current derived in that work are applicable for both the subthreshold and above threshold regions of operation. Although the analytical model is benchmarked against a prototype TMDC transistor, it has some significant limitations like non-scalability due to lumped capacitor network based intrinsic device characteristics. In 2014 Cao et al. [2] presented an analytical I-V model for 2D TMDC FETs as well. The model takes physics of monolayer TMDCs into account and offers a single closed-form expression for all three, i.e. linear, saturation, and subthreshold regions of operation. The authors also incorporated various non-ideal secondary effects like interface traps, mobility degradation, and inefficient doping in the model, although that resulted in current equations having an integral form instead of closed form. In 2015 Najam et al. [3] introduced a surface potential-based low-field drain current compact model for 2D TMDC FET taking dielectric interface traps into account. In that work, the derived drain current model is capable of self-consistently calculating the surface potential of the device and interface trap charge ($Q_{it}$) with the help of an experimentally reported interface trap distribution. In this work, a compact drain current model for monolayer TMDC channel FET has been developed in light of the model developed by Cao et al. [2]. The primary target of this work is to formulate a single drain current equation for all regions of operation. The secondary target is to incorporate the effect of quantum mechanics in the compact drain current model using the appropriate fitting function, which is not present in current literature. In addition to modeling currents in all regions of operation, the compact model developed in this work can capture short channel and non-ideal effects like Drain Induced Barrier Lowering (DIBL), threshold voltage roll-off, mobility degradation, etc. The model invokes gradual channel approximation to simplify the analytical expression and assumes that the electrostatic potential in the channel is limited to quadratic variations only. The model also uses a Field Dependent Mobility Model and considers the E-K diagram obtained from the Density Functional Theory (DFT) to calculate the density of states for monolayer $WSe_2$. The model proposed in this work offers several advantages over Cao's model. In Cao's model [2], one must apply Newton-Raphson approximation to equation 6 (of Cao et al. [2]) to find out the upper and lower limit of $\phi$ which is then used in equation 7 (of Cao et al. [2]) to find out the closed form of current. This was possible because Cao et al. [2] also assumed $\frac{d^2\varphi(x)}{d^2x} = 0$ to simplify the differential equation of the electrostatic potential $\varphi(x)$. In our model, we however assumed $\frac{d^2\varphi(x)}{d^2x} \neq 0$, rather $\frac{d^3\varphi(x)}{d^2x} = 0$, which makes this model sensitive to quadratic variation in $\varphi(x)$. The resulting equation does not have a closed form solution but can be evaluated very easily with numerical integration. Moreover, equation (2) in Cao's work [2] has been simplified to present current in a compact format for all regions of FET operation. Similar simplification is also done in other models too. However, in the model proposed by this work, we have used a linear differential equation with a constant coefficient for electrostatic potential, $\phi(x)$. This helps us to solve the channel potential of a FET device with multiple oxide materials of different dielectric permittivity which is not possible using Cao's model [2] . In fact, the applicability of the proposed model for a FET structure with two different oxide materials in the insulating region has been shown in this work with an example of a dielectric modulated FET type nano-biosensor in the second part (section 7) of this work. In the first part of this work, results obtained from the proposed physically accurate compact model are used for quick characterization of a proposed monolayer $WSe_2$ channel transistor structure. With the help of a gate voltage dependent "Fitting Function" obtained from experimental results of a $MoS_2$ FET (no suitable experimental data on $WSe_2$ FET were available to the authors' best knowledge), this model successfully estimates the transport characteristics and the threshold voltage of the monolayer $WSe_2$ FET, which are also in reasonable agreement with the numerical simulation done in previous work [4]. In the second part of this work, the proposed model has been used to investigate a new type of biosensor which also confirms the versatile applicability of the model. Several TMDC material based FET structures have already been investigated as biosensors in recent literature [5-7]. These monolayer and multilayer TMDC FETs have shown excellent bio-detection capability, especially in the subthreshold mode of operation. However, the impressive performance reported for these biosensors are limited to detection of charged biomolecules submerged in a solution which suffers from well-known Debye screening [8]. New biomolecule detection technique has been proposed in the recent literature [9] which utilizes the dielectric permittivity of the biomolecule instead of charge to circumvent these problems. These biosensors are implemented using dielectric modulated FET (DMFET) where the biomolecules are captured in a cavity formed in the oxide region. Various simulation and analytical model based studies have been reported in the recent literature [10-13] which have proven the potential of this new biosensing technique in a dry environment. This work also investigates the prospect of monolayer TMDC material as a channel in such DMFET biosensor using the compact transport model developed for TMDC MOSFET.

## 2. Device Structure

The MOSFET structure under consideration has a monolayer p-doped $WSe_2$ channel sandwiched between $ZrO_2$ at top and $SiO_2$ at the bottom oxide. Top and bottom gates are made of Pd. The source and drain are highly n-doped regions of monolayer $WSe_2$. The monolayer channel has a thickness, width, and length of 0.65 nm, 10 nm and 20 nm, respectively. Top and bottom oxide thicknesses are 3 nm and 5 nm, respectively.

2D channel is p-doped with an impurity density of 2.2 x $10^{16}$ m⁻² and n-type source/drain doping density are assumed to be in the order of $10^{17}$ m⁻². In our previous work [4], we have performed a quantum mechanical transport simulation on this device structure. A p-FET version of this 2D device structure was first proposed and fabricated by Fang *et al.* [14] which had a 9.2 μm long monolayer WSe₂ channel.

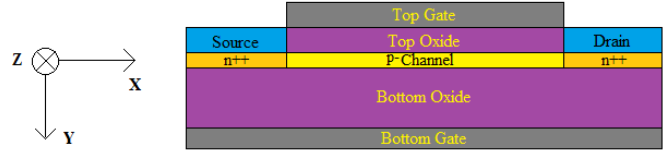

Fig. 1. The MOSFET structure under consideration. It has a 2D material channel sandwiched between the top and bottom oxides and corresponding top and bottom gates. The channel is p-doped.

### 3. Formulation of Electrostatic Potential

A differential system is formulated first to explain the physics and operation of the device as shown in Fig. 2. We have used similar formulation used by Cao *et al.* [2]. Since the channel is very thin, it is reasonable to assume that electrostatic potential $\varphi(x, y)$ in the channel does not change in the direction along the top and bottom gate [2]. So, it is safe to assume that in the channel, potential $\varphi(x, y) \approx \varphi(x)$.

To get the differential system, we need to apply Gauss's Law in the closed box shown in Fig. 2. The box has height $t_{ch}$ (thickness of the 2D channel, ~ 0.65nm), width $W$ and length $x$. From Gauss's Law the relationship between the charge density ($Q$) inside the enclosed box and the electric field outside the enclosed box ($\vec{E}$) can be found as,

$$\oint_s \varepsilon \vec{E} . \vec{s} = Q \qquad (1)$$

where $\varepsilon$ is the dielectric permittivity of the material at each surface ($s$) of the enclosure. Let us assume, the box has an infinitesimal length of $\Delta x$ and charge density of $\Delta Q$. So, Equation (1) becomes,

$$\oint_S \varepsilon \vec{E} . \overrightarrow{ds} = \Delta Q \qquad (2)$$

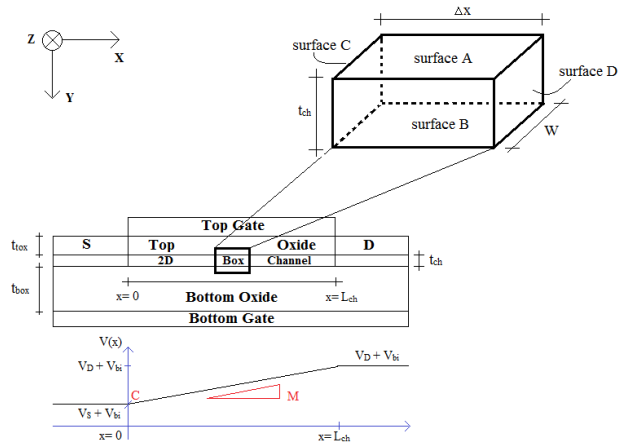

Fig. 2. An infinitesimal box is considered to which Gauss's Law is applied to establish the differential system for the 2D MOSFET. The directions of the surface vectors are outward positive. Approximation of the potential $V(x)$ inside the 2D channel is shown at the bottom of the figure.

### 3.1 Formation of the Differential Equation

From the formulation of Cao $et\ al.$ [2], we get the following differential equation using equation (2),

$$\frac{d^2\varphi(x)}{d^2x} - \varphi(x)\left(\frac{\varepsilon_{tox}}{t_{tox}\varepsilon_{ch}t_{ch}} + \frac{\varepsilon_{box}}{t_{box}\varepsilon_{ch}t_{ch}}\right) + \left(\frac{\varepsilon_{tox}}{t_{tox}\varepsilon_{ch}t_{ch}}V_{Gt}' + \frac{\varepsilon_{box}}{t_{box}\varepsilon_{ch}t_{ch}}V_{Gb}'\right) = \frac{q}{\varepsilon_{ch}t_{ch}}(N_A + n_{2D}(x)) \quad (3)$$

where, $\varepsilon_{tox}$ and $\varepsilon_{box}$ are top and bottom oxide dielectric permittivities, respectively. $t_{tox}$ and $t_{box}$ are top and bottom oxide thicknesses, respectively. $\varepsilon_{ch}$ is the dielectric permittivity of 2D material channel. $N_A$ is the acceptor type dopant concentration per unit area. $n_{2D}(x)$ is the free inversion carrier (electron) concentration. $V_{Gt}'$ and $V_{Gb}'$ are respectively defined as,

$$V_{Gt}' = V_{Gt} - V_{FBt} \quad\quad\quad\quad (4)$$

$$V_{Gb}' = V_{Gb} - V_{FBb} \quad\quad\quad\quad (5)$$

Here, $V_{Gt}$ and $V_{Gb}$ are applied bias voltages at top and bottom gates, respectively and $V_{FBt/b}$ are corresponding flat band voltages. $V_{FBt/b}$ are defined as,

$$V_{FBt} = \phi_{mt} - \phi_{ch} = \phi_{mt} - \left(\chi_{ch} + \frac{E_g}{2q} - \frac{kT}{q}ln\left(\frac{N_A}{n_i}\right)\right) \quad (6)$$

$$V_{FBb} = \phi_{mb} - \phi_{ch} = \phi_{mb} - \left(\chi_{ch} + \frac{E_g}{2q} - \frac{kT}{q}ln\left(\frac{N_A}{n_i}\right)\right) \quad (7)$$

Here, $\phi_{mt}$ and $\phi_{mb}$ are top and bottom metal gate work functions, respectively and $\phi_{ch}$ is the 2D channel material work function. $E_g$, $\chi_{ch}$, and $n_i$ are the bandgap, electron affinity, and intrinsic carrier concentration of the channel material, respectively. $k, T,$ and $q$ are Boltzmann constant, Kelvin temperature (300K), and charge of electron, respectively. Equation (3) can be rewritten as,

$$\frac{d^2\varphi(x)}{d^2x} - R\varphi(x) + G = \frac{q}{\varepsilon_{ch}t_{ch}}(N_A + n_{2D}(x)) \quad\quad\quad (8)$$

Where,

$$G = \frac{\varepsilon_{tox}}{t_{tox}\varepsilon_{ch}t_{ch}}V_{Gt}' + \frac{\varepsilon_{box}}{t_{box}\varepsilon_{ch}t_{ch}}V_{Gb}' \quad\quad\quad (9)$$

$$R = \frac{\varepsilon_{tox}}{t_{tox}\varepsilon_{ch}t_{ch}} + \frac{\varepsilon_{box}}{t_{box}\varepsilon_{ch}t_{ch}} \quad\quad\quad (10)$$

$$n_{2D}(x) = N_{dos}\exp\left(-\frac{E_C(x)-E_F(x)}{kT}\right) = N_{dos}exp\left(\frac{q}{kT}(\varphi(x) - V(x))\right) \quad (11)$$

$E_c(x) = -q\varphi(x)$ is the conduction band profile and $E_F(x) = -qV(x)$ is the quasi Fermi level of the channel. $N_{dos}$ is the effective density of state of the channel material. In equation (11), the effective density of states of WSe$_2$ is represented like that of Cao et al. [2], $N_{dos} = \frac{g_s g_1 kT\ m_1^*}{2\pi\hbar^2} + \frac{g_s g_2 kT\ m_2^*}{2\pi\hbar^2}e^{-\frac{\nabla E_C}{kT}}$. Here, $g_s=$ spin degeneracy, $g_i =$ valley degeneracy, $m_i^* =$ effective mass and $\hbar =$ reduced Plank's constant. The E-K diagram of WSe$_2$ was calculated using Quantum Espresso Software and found that energy difference between the lowest two valleys is ~8meV. So, we had to consider lowest two valleys in the calculation of $N_{dos}$.

Differentiating (8) with respect to $x$ and substituting the value of $\frac{q}{\varepsilon_{ch} t_{ch}} n_{2D}(x)$ from (8) into the result, we get-

$$\frac{d^3\varphi(x)}{d^3x} - R\frac{d\varphi(x)}{dx} = \left[\frac{d^2\varphi(x)}{d^2x} - R\varphi(x) + G - \frac{q}{\varepsilon_{ch} t_{ch}}N_A\right]\left[\frac{q}{kT}\frac{d\varphi(x)}{dx} - \frac{q}{kT}\frac{dV(x)}{dx}\right] \quad (12)$$

The differential equation in (12) cannot be solved for a closed form analytical solution. To simplify the equation, we invoked gradual channel approximation and got $\frac{dV(x)}{dx} \approx 0$ [15]. This assumption is particularly valid for long channel 2D MOSFETs where lateral electric field from drain to source is weaker compared to the vertical electric field from top to bottom gate.

We can further simplify (12) by ignoring higher order variations of $\varphi(x)$ with $x$. As long as the channel is long and drain voltage is low, this assumption is also valid and we get,

$$\frac{d\varphi(x)}{dx}\left[\frac{d^2\varphi(x)}{d^2x} - R\varphi(x) + G - \frac{q}{\varepsilon_{ch} t_{ch}}N_A + R\frac{kT}{q}\right] = 0 \quad (13)$$

Since lateral electric field is non-zero (i.e. $\frac{d\varphi(x)}{dx} \neq 0$) when voltage is applied to the drain, (13) reduces to a linear differential equation with constant co-efficient. The closed form solution of this differential equation is given by,

$$\varphi(x) = C_1 exp\left(\sqrt{R}x\right) + C_2 exp\left(-\sqrt{R}x\right) + \frac{A}{R} \quad (14)$$

Where,

$$A = \frac{kT}{q}R + G - \frac{q}{\varepsilon_{ch} t_{ch}}N_A \quad (15)$$

### 3.2 Evaluating the Constants $C_1$ and $C_2$

For source and drain region, $n_{2d} = N_{D(source)} = N_{D(drain)} = N_{sd}$, where $N_{sd}$ is the n-type source and drain doping concentration per unit area. So, from (11) at $x = 0$ and $x = L_{ch}$ respectively,

$$\varphi(0) = V_S + V_{bi} + \frac{kT}{q}ln\left(\frac{N_{sd}}{N_{dos}}\right) \quad (16)$$

$$\varphi(L_{ch}) = V_D + V_{bi} + \frac{kT}{q}ln\left(\frac{N_{sd}}{N_{dos}}\right) \quad (17)$$

Here, $V_{bi}$ is the built-in potential at the source/drain-channel interface. Using these boundary conditions, we get values of $C_1$ and $C_2$ and (14) becomes,

$$\varphi(x) = \left[ \frac{\left(\varphi(0) - \frac{A}{R}\right) exp\left(-\sqrt{R}L_{ch}\right) - \left(\varphi(L_{ch}) - \frac{A}{R}\right)}{exp\left(-\sqrt{R}L_{ch}\right) - exp\left(\sqrt{R}L_{ch}\right)} \right] exp\left(\sqrt{R}x\right)$$

$$+ \left[ \frac{\left(\varphi(0) - \frac{A}{R}\right) exp\left(\sqrt{R}L_{ch}\right) - \left(\varphi(L_{ch}) - \frac{A}{R}\right)}{exp\left(\sqrt{R}L_{ch}\right) - exp\left(-\sqrt{R}L_{ch}\right)} \right] exp\left(-\sqrt{R}x\right) + \frac{A}{R} \qquad (18)$$

Here, $\frac{A}{R}$ can be evaluated from (10) and (15) as,

$$\frac{A}{R} = \frac{\frac{kT}{q}\left(\frac{\varepsilon_{tox}}{t_{tox}\varepsilon_{ch}t_{ch}} + \frac{\varepsilon_{box}}{t_{box}\varepsilon_{ch}t_{ch}}\right) + \left(\frac{\varepsilon_{tox}}{t_{tox}\varepsilon_{ch}t_{ch}}V'_{Gt} + \frac{\varepsilon_{box}}{t_{box}\varepsilon_{ch}t_{ch}}V'_{Gb}\right) - \frac{q}{\varepsilon_{ch}t_{ch}}N_A}{\frac{\varepsilon_{tox}}{t_{tox}\varepsilon_{ch}t_{ch}} + \frac{\varepsilon_{box}}{t_{box}\varepsilon_{ch}t_{ch}}} \qquad (19)$$

## 4. Drain Current Modeling

The carrier transport is governed by the drift-diffusion equation [2, 16] as described by,

$$I_x(x) = qWn_{2D}(x)\mu_n(x)\frac{dV(x)}{dx} \qquad (20)$$

Here, $\mu_n(x)$ is the channel electron mobility. In this stage, we need an estimation of $V(x)$ in terms of $x$ to calculate the current. The most simplified approximation can be a linear profile of $V(x)$ as described by the equation,

$$V(x) = Mx + C \qquad (21)$$

From Fig. 2 the constants $M$ and $C$ can be evaluated as,

$$M = \frac{(V_D - V_S)}{L_{ch}} \qquad (22)$$

$$C = V_S + V_{bi} \qquad (23)$$

The value of $M$ is consistent with the gradual channel approximation. With longer $L_{ch}$ and lower $V_D$, $M$ gets smaller and dependence of $V(x)$ on $x$ diminishes to give $\frac{dV(x)}{dx} \approx 0$. However, in the value of $M$ and $C$ from (22) and (23), gate voltage dependence of quasi Fermi level is missing. To incorporate the effect of gate voltage, an empirical fitting function $F(V_G)$ can be considered with $C$. So, the final form of $C$ becomes,

$$C = V_S + V_{bi} + F(V_G) \qquad (24)$$

Assuming a uniform drain current $I_{DS}$ and uniform field dependent mobility $\mu_n$ throughout the channel and integrating (20) with respect to $x$ from $x = 0$ to $x = L_{ch}$ and substituting value of $n_{2D}(x)$, we get-

$$I_{DS} = \frac{qWMN_{dos}}{L_{ch}}\mu_n \int_{x=0}^{x=L_{ch}} exp\left(\frac{q}{kT}\left(\varphi(x) - V(x)\right)\right)dx \qquad (25)$$

The lateral electric field ($E_{||}$) dependence of the mobility ($\mu_n$) will come from a standard mobility model as used in ATLAS [17],

$$\mu_n = \frac{\mu_{n0}}{\left[1 + \left[\frac{\mu_{n0} E_{||}}{v_{satn}}\right]^{\beta n}\right]^{\frac{1}{\beta n}}} \qquad (26)$$

where,

$\mu_{n0}$ = Low field mobility

$v_{satn}$ = Electron saturation velocity in the electric field

$\beta_n$ = Fitting parameter

$E_{||}$ = Lateral electric field from drain to source = $\frac{V_D - V_S}{L_{ch}}$

Substituting values of $\varphi(x)$, $V(x)$, and $\mu_n$ into (25), we get the final expression of the drain current per unit channel width as,

$$I_{DS} = \frac{qMN_{dos}}{L_{ch}} \frac{\mu_{n0}}{\left[1 + \left[\frac{\mu_{n0} E_{||}}{v_{satn}}\right]^{\beta n}\right]^{\frac{1}{\beta n}}} \int_{x=0}^{x=L_{ch}} exp\left(\frac{q}{kT}\left(C_1 exp(\sqrt{R}x) + C_2 exp(-\sqrt{R}x) + \frac{A}{R} - Mx - V_S - V_{bi} - F(V_G)\right)\right) dx \quad (27)$$

Although the integral in (27) does not have a closed form solution, it can be very easily evaluated numerically to get the drain current.

## 5. Model Verification

Due to the unavailability of experimental transport characteristics of sub-1µm channel monolayer WSe$_2$ FET, we had to choose an experimental monolayer MoS$_2$ FET for the benchmarking purpose. The compact transport characteristics developed in the previous section is compared with the experimental results of monolayer MoS$_2$ FET published by Liu *et al.* [18]. Precisely same sets of the device and material parameters are used to make a valid comparison. Since monolayer MoS$_2$ and WSe$_2$ have similar crystal structure and material properties, the fitting function $F(V_G)$ obtained from the experimental results are assumed to be valid for the monolayer WSe$_2$ FET as well. Applying this fitting function to the WSe$_2$ FET described in section 2, we get its transport characteristics which reasonably match with the self-consistent simulation results from our previous work [4] on monolayer WSe$_2$ FET.

Fig. 3(a) shows the I-V characteristics of the monolayer MoS$_2$ device obtained from equation (27) along with the experimental results reported by Liu *et al.* [18]. To match the characteristics, fitting function $F(V_G)$, low field mobility ($\mu_{n0}$) and $\beta_n$ have been configured as shown in Table I.

TABLE I
FITTING PARAMETERS FOR THE COMPACT DRAIN CURRENT MODEL

| Parameter | Symbol | Value |
|---|---|---|
| Fitting Function | $F(V_G)$ | $a(V_{Gt} + b)^c + d$ |
| Low Field Mobility | $\mu_{n0}$ | 40 x 10$^{-4}$ m$^2$/V.s |
| Fitting Constant | $\beta_n$ | 2.5 |
| Electron Saturation Velocity | $v_{satn}$ | 2.2 x 10$^5$ ms$^{-1}$ |

The fitting function $F(V_G)$ has parameters $a = -0.4609, b = 0.25, c = 0.2099$ and $d = 0.3527$ for benchmarking with the experimental data in Fig. 3(a). However, considering a marginal change of a parameter (c=0.3397), we found that the proposed model matches with the self-consistent simulation results from our previous work [4] too. Fig. 3(b) shows the transfer characteristics of the monolayer WSe$_2$ device obtained from (27) along with the transfer characteristics from the self-consistent simulation [4].

## 6. Results and Discussions

In this section, we use the fitting functions obtained from the previous section to study the monolayer WSe₂ FET described in section 2.

### 6.1 Material Parameters Used

TABLE II
MATERIAL PARAMETERS FOR THE DEVICE

| Parameter | Symbol | Value | Unit |
|---|---|---|---|
| Monolayer WSe₂ Electron Effective Mass | $m^*$ | $0.33 m_e$ | kg |
| Monolayer WSe₂ Dielectric Permittivity | $\varepsilon_{ch}$ | $5.2\varepsilon_o$ | Fm$^{-1}$ |
| Monolayer WSe₂ Bandgap | $E_g$ | 1.6 | eV |
| Monolayer WSe₂ Electron Affinity | $\chi_{ch}$ | 3.9 | eV |
| Pd Work Function | $\Phi_m$ | 5.1 | eV |
| ZrO₂ Dielectric Permittivity | $\varepsilon_{tox}$ | $12.5\varepsilon_o$ | Fm$^{-1}$ |
| SiO₂ Dielectric Permittivity | $\varepsilon_{box}$ | $3.9\varepsilon_o$ | Fm$^{-1}$ |

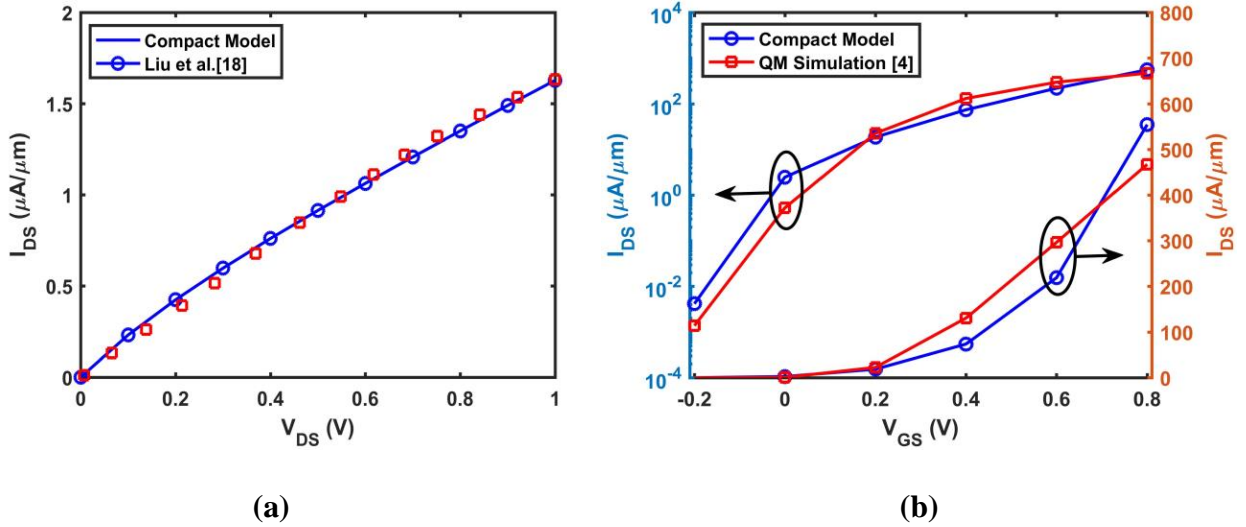

**(a)** **(b)**

Fig. 3. Drain current from the compact model is matched with (a) experimental monolayer MoS₂ FET results [18] for $V_{GS}$ = 1V (b) Quantum mechanical transport simulated WSe₂ FET's transfer characteristics [4] for $V_{DS}$ = 0.8V. Plots are made by using appropriate material and fitting function in both real (right y-axis) and semi-log (left y-axis) scale.

## 6.2 Transport Characteristics of the Device

Fig. 4(a) and 4(b) displays the channel potential $\varphi(x)$ under different top gate bias conditions. The bottom gate voltage is kept at zero. Fig. 4(b) demonstrates the effect of lateral electric field by showing minor change in the channel potential near the drain end despite of the same gate voltage. Fig. 5(a) displays the output characteristics ($I_{DS}$-$V_{DS}$) of the device under different top gate voltages.

## 6.3 Threshold Voltage and Performance Parameters Extraction

Figure 5(b) presents the $I_{DS} - V_{GS}$ characteristics in semi-logarithmic scale. DIBL and Subthreshold Swing (SS) are calculated as 40 mV/V (evaluated between $V_{DS} = 0.1$V and $V_{DS} = 2$V) and 72 mV/Dec (at $V_{DS} = 0.1$V), respectively. From the figure, threshold voltages were calculated as 0.272 V, 0.260 V and 0.256 V respectively for 0.1 V, 1 V and 2 V drain biases. This dependence of threshold voltage on drain bias arises from the effect of lateral electric field in the channel due to shorter channel length.

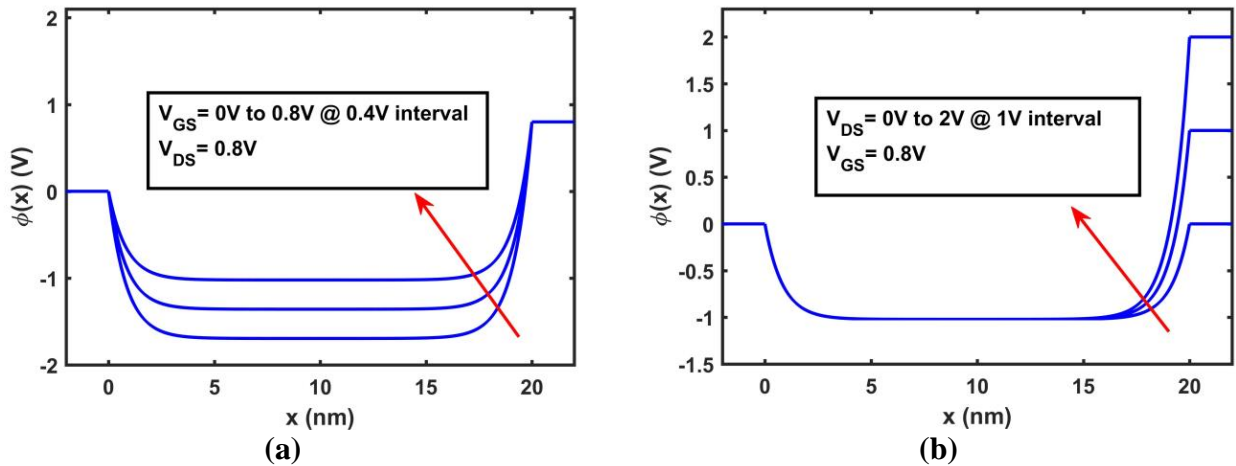

Fig. 4. Channel potential $\varphi(x)$ under different (a) top gate bias conditions for $V_{DS} = 0.8$V and (b) drain bias conditions for $V_{GS} = 0.8$V. In both cases bottom gate voltage is considered as zero.

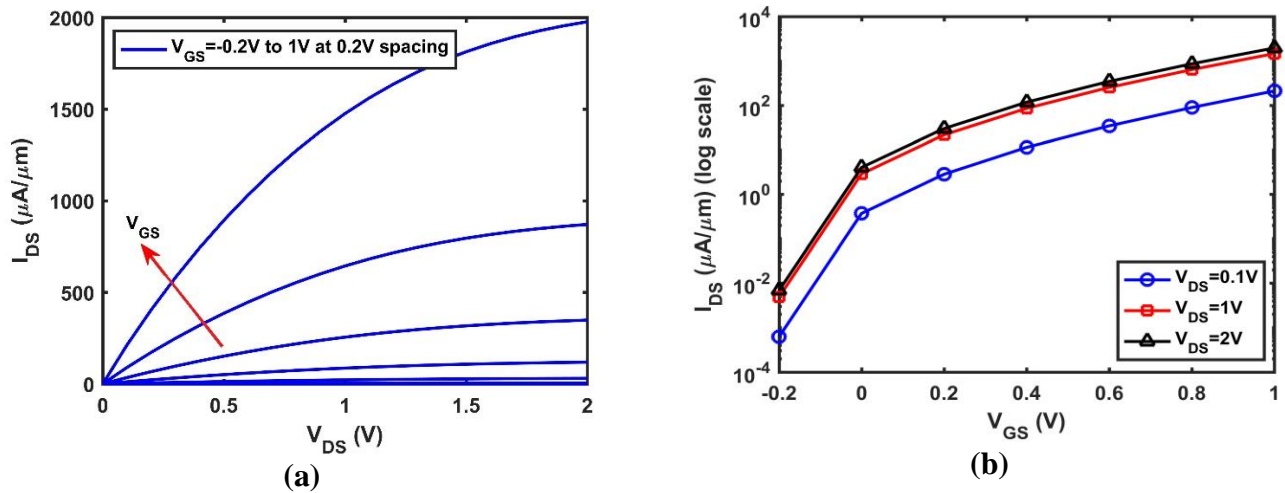

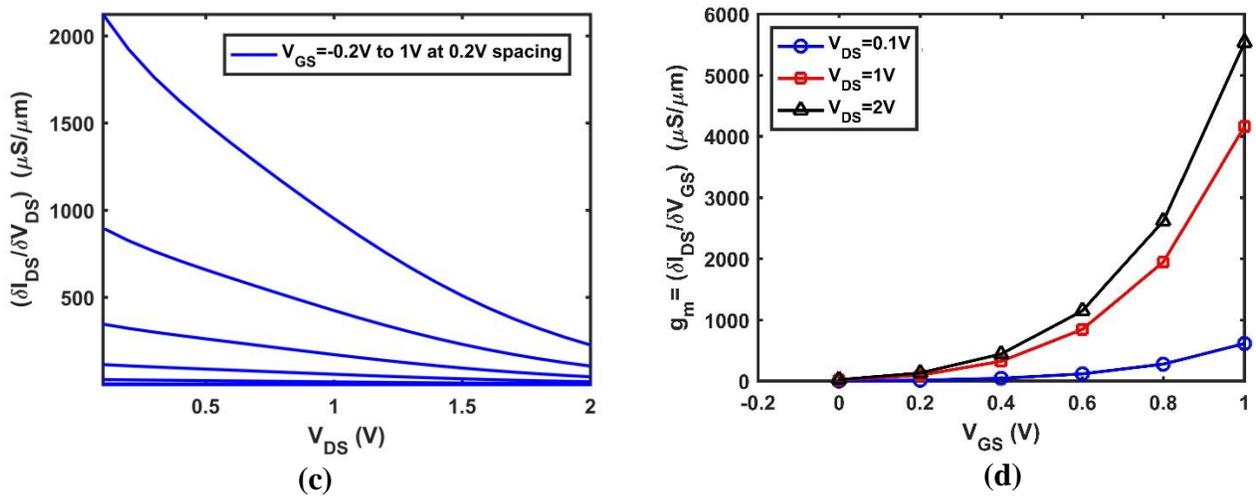

**(c)**  **(d)**

Fig. 5. (a) Output characteristics ($I_{DS}$-$V_{DS}$) and (b) transfer characteristics ($I_{DS}$-$V_{GS}$) (c) channel conductance ($\frac{\delta I_{DS}}{\delta V_{DS}}$) and (d) transconductance ($\frac{\delta I_{DS}}{\delta V_{GS}}$) of the device under different bias voltages. Bottom gate is fixed at zero.

Figure 5(c) and 5(d) present the channel conductance, $g_{ch}$ ($\frac{\delta I_{DS}}{\delta V_{DS}}$) and transconductance, $g_m$ ($\frac{\delta I_{DS}}{\delta V_{GS}}$) of the device under different bias voltages respectively. From the output and transfer characteristics of the proposed monolayer p-WSe$_2$ channel MOSFET, we can consider it as a potential candidate for the next generation high-speed, low power applications.

## 7. DMFET Biosensor

The biosensor studied in this work is shown in Fig. 6(a). It consists of a top and bottom oxide with top oxide thickness equal to the height of the biomolecule. The channel material is a single layer MoS$_2$. The channel has a p-type doping with a doping density of 2.2x10$^{16}$ m$^{-2}$. Source and drains are highly doped with an n-type impurity with a doping density of 3.25x10$^{17}$ m$^{-2}$. Therefore, it results in an inversion mode dielectrically modulated FET. The detailed derivation of the proposed model for biosensor application has been provided in supplementary document S1.

TABLE III
MATERIAL AND DEVICE PARAMETERS FOR MONOLAYER MoS$_2$ DMFET

| Parameter | Symbol | Value | Unit |
|---|---|---|---|
| Channel Thickness | $t_{ch}$ | 0.7 | nm |
| Biomolecule Length | $L_{bio}$ | 30 | nm |
| Top Oxide Length | $L_{tox}$ | 30 | nm |
| Channel Length | $L_{ch}$ | 60 | nm |

| Channel Doping | $N_A$ | $2.2 \times 10^{16}$ | $m^{-2}$ |
|---|---|---|---|
| Source/Drain Doping | $N_D$ | $3.25 \times 10^{17}$ | $m^{-2}$ |
| Top Oxide Thickness (SiO₂) | $t_{tox}$ | 10 | nm |
| Biomolecule Thickness | $t_{bio}$ | 10 | nm |
| Bottom Oxide Thickness (SiO₂) | $t_{box}$ | 10 | nm |
| Monolayer MoS₂ Electron Effective Mass | $m^*$ | $0.56\, m_e$ | kg |
| Monolayer MoS₂ Dielectric Permittivity | $\varepsilon_{ch}$ | $8.29\, \varepsilon_o$ | $Fm^{-1}$ |
| SiO₂ Dielectric Permittivity | $\varepsilon_{tox}\, \&\, \varepsilon_{box}$ | $3.9\, \varepsilon_o$ | $Fm^{-1}$ |
| Biomolecule Dielectric Permittivity | $\epsilon_{bio}$ | $3{\sim}9\, \varepsilon_o$ | $Fm^{-1}$ |

## 7.1 Result and Discussion

As the dielectric permittivity of biomolecules in the cavity region changes, the potential profile also changes according to the relative dielectric constant of the biomolecule. Fig. 6(b) shows the surface potential when the dielectric permittivity of the cavity region is identical to that of top and bottom oxide, which is SiO₂ in both region. In this specific case, the surface potential does not show any deformation in the oxide cavity interface. The whole top oxide and cavity region act as a uniform material as far as the dielectric permittivity of these two regions is concerned. However, for any other biomolecule with a dielectric permittivity different from that of the oxide region, change in surface potential is observed at the oxide-cavity interface. As can be seen from Fig. 7, a higher dielectric permittivity of biomolecule will cause the surface potential under the cavity to shift upward compared to the rest of the oxide region. The extent of such shift in channel electrostatic potential will depend on the difference in relative dielectric constant between the cavity region and top oxide. A significant upward shift in surface potential under the cavity will cause an increase in inversion carrier density inside the channel. The increase inversion carrier density modulates the channel conductivity, which ultimately results in higher drain current as evident from Fig. 8. For assessing the sensing performance of the proposed device, we used sensitivity metric as follows

$$S = \frac{\Delta I_{DS}}{I_{DS}} = \frac{|I_{DS}(Bio) - I_{DS}(Air)|}{I_{DS}(Air)} \qquad (28)$$

Fig. 9(a) shows the percentage change in drain current for the different relative dielectric constant of the biomolecule with cavity height as a parameter. It can be observed from Fig. 9(a) that as the dielectric constant of the biomolecule increases, the sensitivity also increases with an upper level of around 45%. This change in sensitivity should be significant enough to be measured considering the signal fluctuation caused by background noise. Moreover, an increase in cavity height also increases the sensitivity. Fig. 9(b) shows the percentage change in drain current with the relative dielectric constant of the biomolecule for different cavity length. From equation (27), it can be seen that an increase in cavity length will decrease the current. A reduction in current also reduces the sensitivity as seen from Fig. 9(b). Therefore, to optimize the sensitivity of the proposed device, the sensor should be designed in such a way so that

a) The difference in relative dielectric constant between oxide and cavity region is maximum.

b) The area of the cavity region is adequate to cause considerable change in device's electrical response, i.e., drain current.

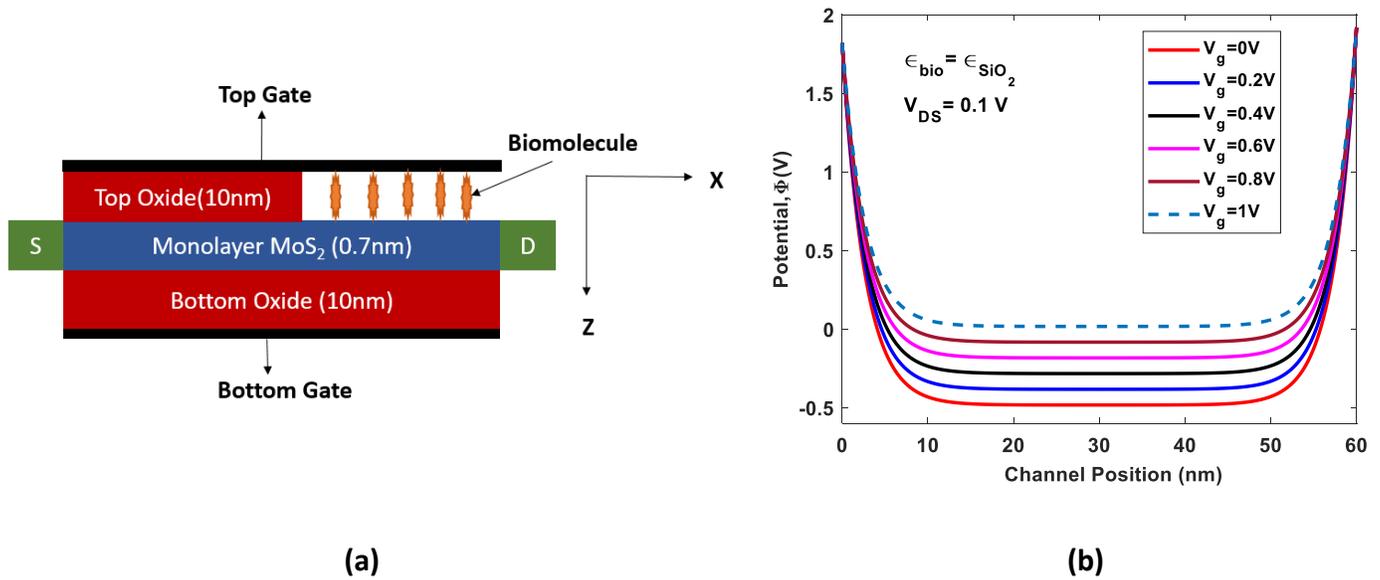

**(a)**                                                                 **(b)**

Fig.6. (a) The DMFET biosensor structure under consideration. Monolayer MoS$_2$ is used as channel material sandwiched between the top and bottom oxides. The channel is p-doped while the source and drain are highly n-doped regions of the same 2D material. A cavity region in the top oxide with a different dielectric constant is used to simulate the effect of the biomolecule on the conductivity of the FET. Top and bottom oxides are identical regarding the type of material and dimensions. (b) Channel potential profile $\varphi(x)$ when biomolecule dielectric permittivity is equal to that of top oxide.

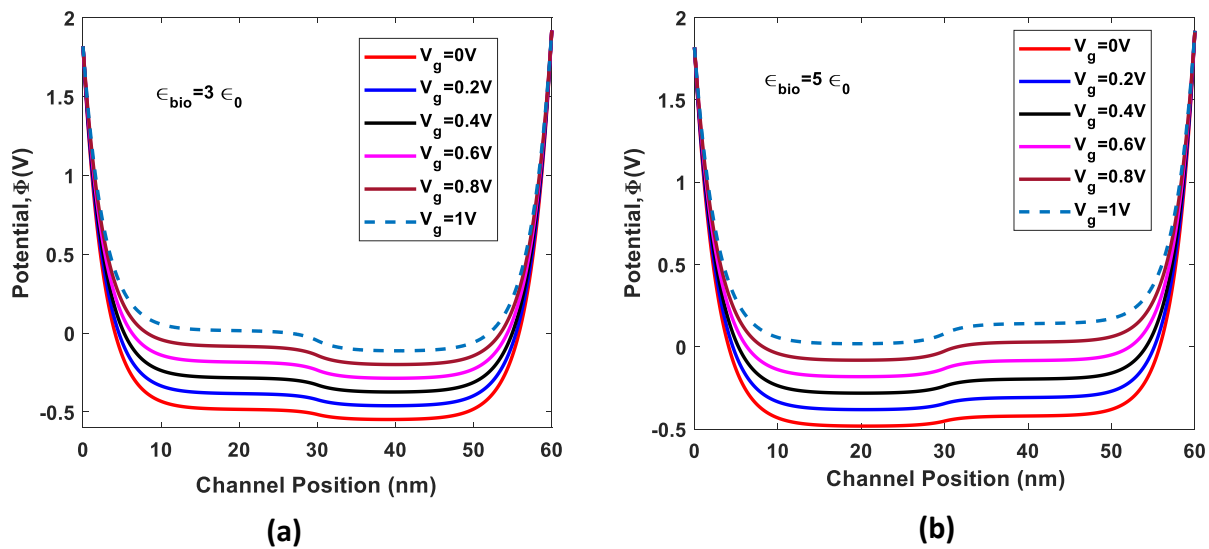

**(a)**                                                                 **(b)**

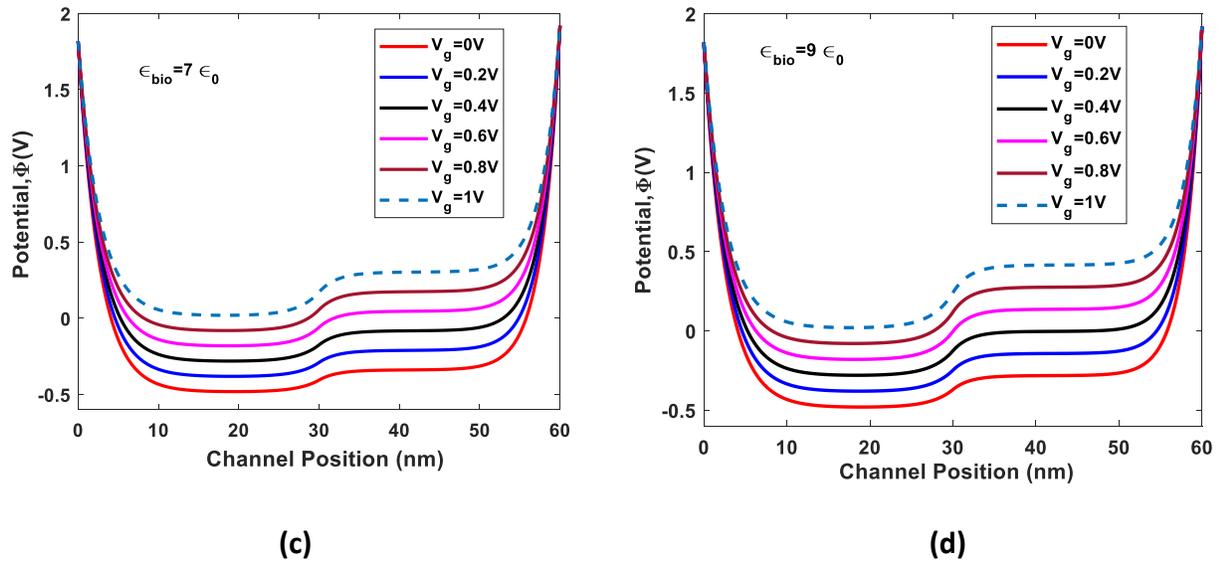

**(c)**                                            **(d)**

Fig.7. Channel potential profile $\varphi(x)$ when the cavity is filled with biomolecules of different dielectric permittivity for different gate voltage , $V_g$ at $V_{DS}$=0.1V.

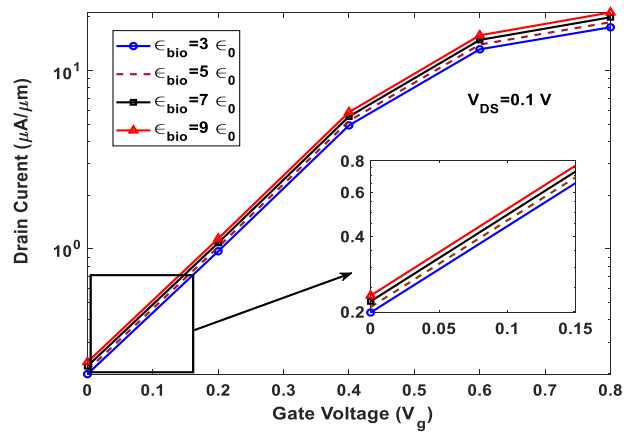

Fig.8. Drain Current for different dielectric permittivity. Inset shows zoomed view of the subthreshold region.

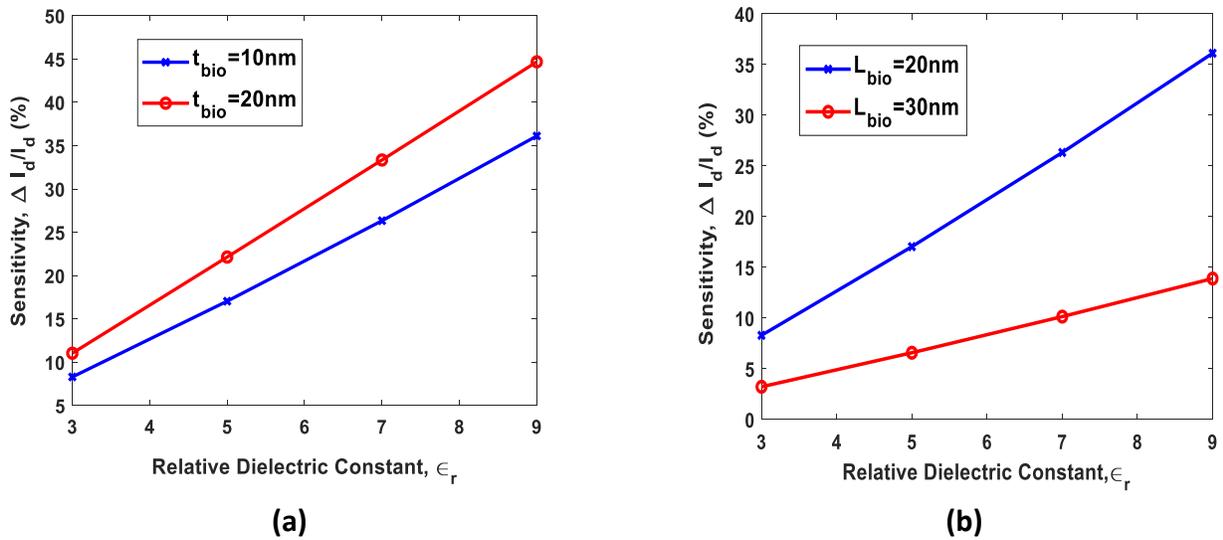

**(a)**                                            **(b)**

Fig.9. (a) DMFET sensitivity for different biomolecule dielectric permittivity with oxide thickness as a parameter (b) Percentage change in drain current for different dielectric permittivity with cavity length as a parameter.

## 8. Conclusion

In this paper, a physically based accurate compact model of monolayer TMDC MOSFET and DMFET biosensor has been developed through a comprehensive analytical study by solving Poisson's equation for the electrostatic potential in the channel region and the drift-diffusion equation for the drain current using the gradual channel and quadratic electrostatic potential approximation. The results from the model have shown excellent agreement with the experimental and self-consistent simulation results for MOSFETs available in the current literature. Potential profiles, output, and transfer characteristics of monolayer $WSe_2$ MOSFET have been calculated from the compact model, and the results confirm excellent ON and OFF state performances of monolayer $WSe_2$ MOSFET. The compact model also is used to design and explain the operation of an electronic biosensor working on the principle of dielectric permittivity modulation. The DMFET biosensor channel potential and transfer characteristics obtained from the model show a detectable change in sensor output with the variation of dielectric permittivity in the cavity region. The results indicate that the sensitivity of DMFET biosensor can be optimized through proper selection of the cavity dimension and oxide materials.

## References

[1] D. Jiménez. "*Drift-diffusion model for single layer transition metal dichalcogenide field-effect transistors.*" Applied Physics Letters, vol. 101, no. 24, p. 243501, 2012.

[2] W. Cao, J. Kang, W. Liu, and K. Banerjee. "*A compact current–voltage model for 2D semiconductor based field-effect transistors considering interface traps, mobility degradation, and inefficient doping effect.*" IEEE Transactions on Electron Devices, vol. 61, no. 12, pp. 4282-4290, 2014.

[3] F. Najam, M. L. P. Tan, R. Ismail, and Y. S. Yu. "*Two-dimensional (2D) transition metal dichalcogenide semiconductor field-effect transistors: the interface trap density extraction and compact model.*" Semiconductor Science and Technology, vol. 30, no. 7, p. 075010, 2015.

[4] S. U. Z. Khan and Q. D. M. Khosru. "*Quantum Mechanical Electrostatics and Transport Simulation and Performance Evaluation of Short Channel Monolayer $WSe_2$ Field Effect Transistor.*" ECS Transactions, vol. 66, no. 14, pp. 11-18, 2015.

[5] A. Shadman, E. Rahman, and Q.D. M. Khosru, "*Monolayer $MoS_2$ and $WSe_2$ Double Gate Field Effect Transistor as Super Nernst pH sensor and Nanobiosensor.*" Sensing and Bio-Sensing Research, vol. 11, no. 1, pp. 45–51, 2016.

[6] K. Datta, A. Shadman, E. Rahman, and Q. D. M. Khosru, "*Trilayer TMDC Heterostructures for MOSFETs and Nanobiosensors.*" Journal of Electronic Materials, vol. 46, no. 2, pp. 1248–1260, 2017.

[7] D. Sarkar, W. Liu, X. Xie, A. C. Anselmo, S. Mitragotri, and K. Banerjee, "*$MoS_2$ Field-Effect Transistor for Next-Generation Label-Free Biosensors.*" ACS Nano, vol. 8, no. 4, pp 3992–4003, 2014.

[8] E. Stern, R. Wagner, F. J. Sigworth, R. Breaker, T. M. Fahmy, and M. A. Reed, "*Importance of the debye screening length on nanowire field effect transistor sensors,*" Nano Letters, vol. 7, no. 11, pp. 3405–3409, 2007.

[9] H. Im, X.-J. Huang, B. Gu, and Y.-K. Choi, "*A dielectric-modulated field-effect transistor for biosensing.*" Nature Nanotechnology, vol. 2, pp. 430–434, 2007.

[10] M. Verma, S.Tirkey, S.Yadav, D. Sharma and D. S. Yadav, "*Performance Assessment of A Novel Vertical Dielectrically Modulated TFET-Based Biosensor.*" IEEE Transactions on Electron Devices, vol. 64, no. 9, pp. 3841 – 3848, 2017.

[11] P. Venkatesh, K. Nigam, S. Pandey, D. Sharma and P.N.Kondekar, "*A dielectrically modulated electrically doped tunnel FET for application of label free biosensor.*" Superlattices and Microstructures, vol.109, pp. 470-479, 2017.

[12] A. Shadman, E. Rahman, and Q.D. M. Khosru, "*Quantum ballistic analysis of transition metal dichalcogenides based double gate junctionless field effect transistor and its application in nano-biosensor*." Superlattices and Microstructures, vol.111, pp. 414-422, 2017.

[13] E. Rahman, A. Shadman, and Q.D. M. Khosru, "*Effect of biomolecule position and fill in factor on sensitivity of a Dielectric Modulated Double Gate Junctionless MOSFET biosensor*." Sensing and Bio-Sensing Research, vol. 13, pp. 49-54, 2017.

[14] H. Fang, S. Chuang, T. C. Chang, K. Takei, T. Takahashi, and A. Javey. "*High-performance single layered $WSe_2$ p-FETs with chemically doped contacts*." Nano Letters, vol. 12, no. 7, pp. 3788-3792, 2012.

[15] D. Jimenez, E. Miranda, and A. Godoy. "*Analytic model for the surface potential and drain current in negative capacitance field-effect transistors.*" IEEE Transactions on Electron Devices, vol. 57, no. 10, pp. 2405-2409, 2010.

[16] D. Neamen. Semiconductor physics and devices. 3$^{rd}$ Ed., New York: McGraw-Hill, 2002, pp. 494.

[17] SILVACO Simulation Standards. [Online] Available:http://www. silvaco.com/tech_lib_TCAD/simulation-standard/2000/feb/a2/ a2.html

[18] H. Liu, M. Si, S. Najmaei, A. T. Neal, Y. Du, P. M. Ajayan, J. Lou, and P. D. Ye. "*Dual-gate MOSFETs on monolayer CVD $MoS_2$ films.*" Device Research Conference (DRC), pp. 163-164, 2013.